# Observation of the Triplet Spin-Valve Effect in a Superconductor-Ferromagnet Heterostructure

V. Zdravkov[1,2], J. Kehrle[1], G. Obermeier[1], D. Lenk[1], H.-A. Krug von Nidda[1], C. Müller[1], M. Yu. Kupriyanov[3], A. S. Sidorenko[2], S. Horn[1], R. Tidecks[1], L. R. Tagirov[1,4]

[1]*Institut für Physik, Universität Augsburg, D-86159 Augsburg, Germany*

[2]*Institute of Electronic Engineering and Nanotechnologies ASM, Kishinev MD-2028, Moldova*

[3]*Skobeltsyn Institute of Nuclear Physics, Moscow State University, Moscow 119992, Russia*

[4]*Solid State Physics Department, Kazan Federal University, 420008 Kazan, Russia*

The theory of superconductor-ferromagnet (S-F) heterostructures with two ferromagnetic layers predicts the generation of a long-range, odd-in-frequency triplet pairing at non-collinear alignment (NCA) of the magnetizations of the F-layers. This triplet pairing has been detected in a Nb/$Cu_{41}Ni_{59}$/nc-Nb/Co/$CoO_x$ spin-valve type proximity effect heterostructure, in which a very thin Nb film between the F-layers serves as a normal conducting (nc) spacer. The resistance of the sample as a function of an external magnetic field shows that for not too high fields the system is superconducting at a collinear alignment of the $Cu_{41}Ni_{59}$ and Co layer magnetic moments, but switches to the normal conducting state at a NCA configuration. This indicates that the superconducting transition temperature $T_c$ for NCA is lower than the fixed measuring temperature. The existence of a minimum $T_c$, at the NCA regime below that one for parallel or antiparallel alignments of the F-layer magnetic moments, is consistent with the theoretical prediction of a singlet superconductivity suppression by the long-range triplet pairing generation.

An odd-in-frequency triplet pairing generation in singlet superconductor/ferromagnet thin-film heterostructures was predicted theoretically [1-3]. At least two ferromagnetic layers ($F_1$,$F_2$) with a non-collinear alignment of their magnetizations, are required to couple the conventional opposite-spin singlet s-wave pairing channel with the unconventional, odd-triplet s-wave pairing channel. The latter one is of extraordinary long range in F layers [1,2,4], because the magnetized conduction band of a ferromagnetic metal serves as an eigenmedia supporting the equal-spin pairing.

Intense activities followed to formulate optimal conditions and realize experimental schemes for the generation and detection of this odd-triplet pairing utilizing the Josephson effect [5-14]. The observation of a current crossing a weak link of ferromagnetic material with a thickness much exceeding the penetration length for singlet-paired electrons [5-11] indicated a triplet contribution to the Josephson current.

In superconductor-ferromagnet proximity-type experiments, also the odd-triplet pairing was considered [15,16]. In a recent paper [17] a deep absolute minimum of the superconducting transition temperature, $T_c$, due to the odd-triplet component generation was predicted for a S/$F_1$/$F_2$ superconducting spin-valve heterostructure near the crossed configuration of the magnetic moments of the adjacent $F_1$ and $F_2$ layers. The aim of the present work is to realize this odd-triplet pairing induced spin-valve effect experimentally.

An S/$F_1$/N/$F_2$/AF spin-valve heterostructure (Fig. 1) was used. Here, S is a singlet superconductor (Nb), $F_1$ and $F_2$ are metallic ferromagnet layers ($Cu_{41}Ni_{59}$ alloy and Co), N is a normal conducting (nc) metal spacer (very thin Nb, below the critical thickness [18]), and AF denotes an insulating antiferromagnet ($CoO_x$), to exchange bias the magnetic moment of the $F_2$ layer.

The equilibrium magnetization of the $Cu_{41}Ni_{59}$ alloy is perpendicular to the layer plane [19,20], whereas for thin Co it lies in the film plane [21]. Then, with an in-plane external magnetic field one could control the magnetic configuration of the system from a parallel alignment, through a crossed one, towards an antiparallel alignment of the F-layer magnetic moments (see the sketch in Fig. 1 and measurements below).

To get samples with different thicknesses of the $Cu_{41}Ni_{59}$ alloy, a wedge-shaped layer was deposited by magnetron sputtering as described in Refs [18,22]. The resulting specimen, Nb/$Cu_{41}Ni_{59}$-wedge/nc-Nb/Co/$CoO_x$, was cut perpendicular to the CuNi thickness gradient on 25 stripes numbered from #1 to #25 starting from the thick side. A pilot series of Nb/$Cu_{41}Ni_{59}$/Si-cap S/F bilayers for magnetoresistance measurements, and a four-wedge ($Cu_{41}Ni_{59}$-wedge/Si)×4 sample reference series for magnetic measurements were fabricated by the same technique.

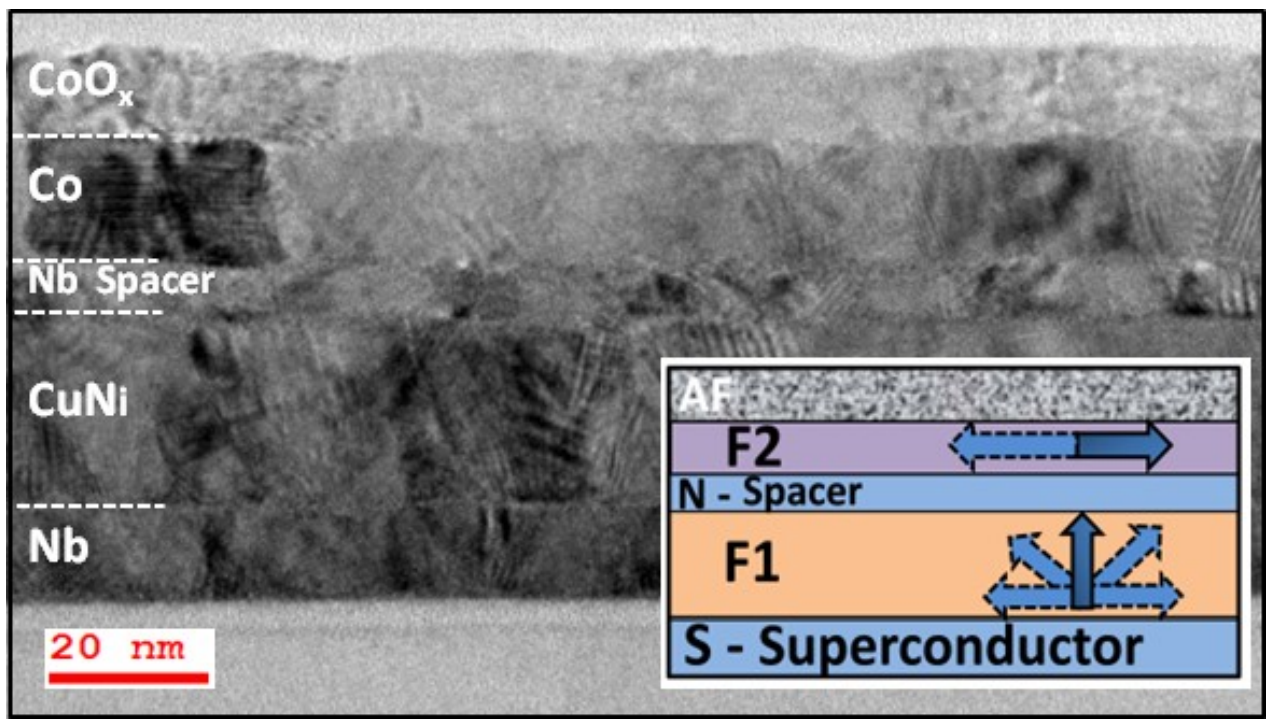

Fig. 1. The Nb/$Cu_{41}Ni_{59}$/nc-Nb/Co/$CoO_x$ sample cross section (transmission electron microscope (TEM) image of sample $SF_1NF_2$-AF1#5 and sketch). Arrows in the sketch indicate possible directions of the layers magnetic moments.

We first measured hysteresis loops of the reference ($Cu_{41}Ni_{59}$/Si)×4 samples in directions perpendicular to the sample plane and then, in-plane, parallel and perpendicular to the initial CuNi layer gradient (inset in Fig. 2b) using a superconducting quantum interference device (SQUID) magnetometer. The out-of-plane hysteresis loop (red) clearly shows easy-axis behavior (larger coercitivity and squareness compared to in-plane loops (blue and black)). The in-plane semi-easy axis was determined as crossed to the wedge gradient direction.

The desired sequence of magnetic configurations in the system was passed applying a magnetic field along the in-plane semi-easy axis of the $Cu_{41}Ni_{59}$ layer, which was simultaneously the easy axis of the Co film. The samples were cooled down at 10 kOe, then the magnetic hysteresis loops were recorded by a SQUID magnetometer in the field range ± 4 kOe. Results of samples $SF_1NF_2$-AF1#1 and #16 are shown in Fig. 2a and the inset, respectively.

For sample #1 (thickest $Cu_{41}Ni_{59}$ alloy layer) the $Cu_{41}Ni_{59}$ and Co layer signal could be separated according to [23] (Fig. 2b), which shows a clear exchange bias of $H_{bias} \approx 940$ Oe due to the antiferromagnetic $CoO_x$. Resulting magnetic configurations are indicated by pictograms. Upon sweeping the field from the positive saturated (PS) configuration at +4 kOe towards the negative saturated (NS) configuration (from – 1.550 Oe to – 4 kOe), the sample passes through the state with crossed (CR) magnetic moments at approximately – 250 Oe, and the (almost) antiparallel alignment (APA) of the Co and $Cu_{41}Ni_{59}$ magnetic moments in the range from – 250 to – 1500 Oe. A similar sequence follows when sweeping the field in the reverse direction. The pilot S/F bilayers behave similar to the $Cu_{41}Ni_{59}$ layer in Fig. 2b (red).

Resistance measurements were performed using the standard DC four-probe method with sensing current 10 μA (polarity alternated to eliminate thermoelectric voltages), flowing parallel to the magnetic field. Prior to the measurements the samples were cooled at 30 kOe in a field applied parallel to the in-plane semi-easy axis as in the magnetic measurements. A set of resistance-temperature, $R(T)$, curves recorded at different magnetic fields $H$ in this direction are given in Fig. 3a.

The magnetoresistance (MR) measurements at $T_1 \approx 3.80$ K, well above the onset of the superconducting (SC) transition at zero field (midpoint $T_c = 3.566$ K) are shown in Fig. 3b. Weak downward peaks of anisotropic magnetoresistance (AMR) coinciding with the $Cu_{41}Ni_{59}$ layer coercive fields. These MR results are consistent with an intrinsic magnetization of the $Cu_{41}Ni_{59}$ layer perpendicular, and that one of the Co layer parallel, to the film plane and to the current.

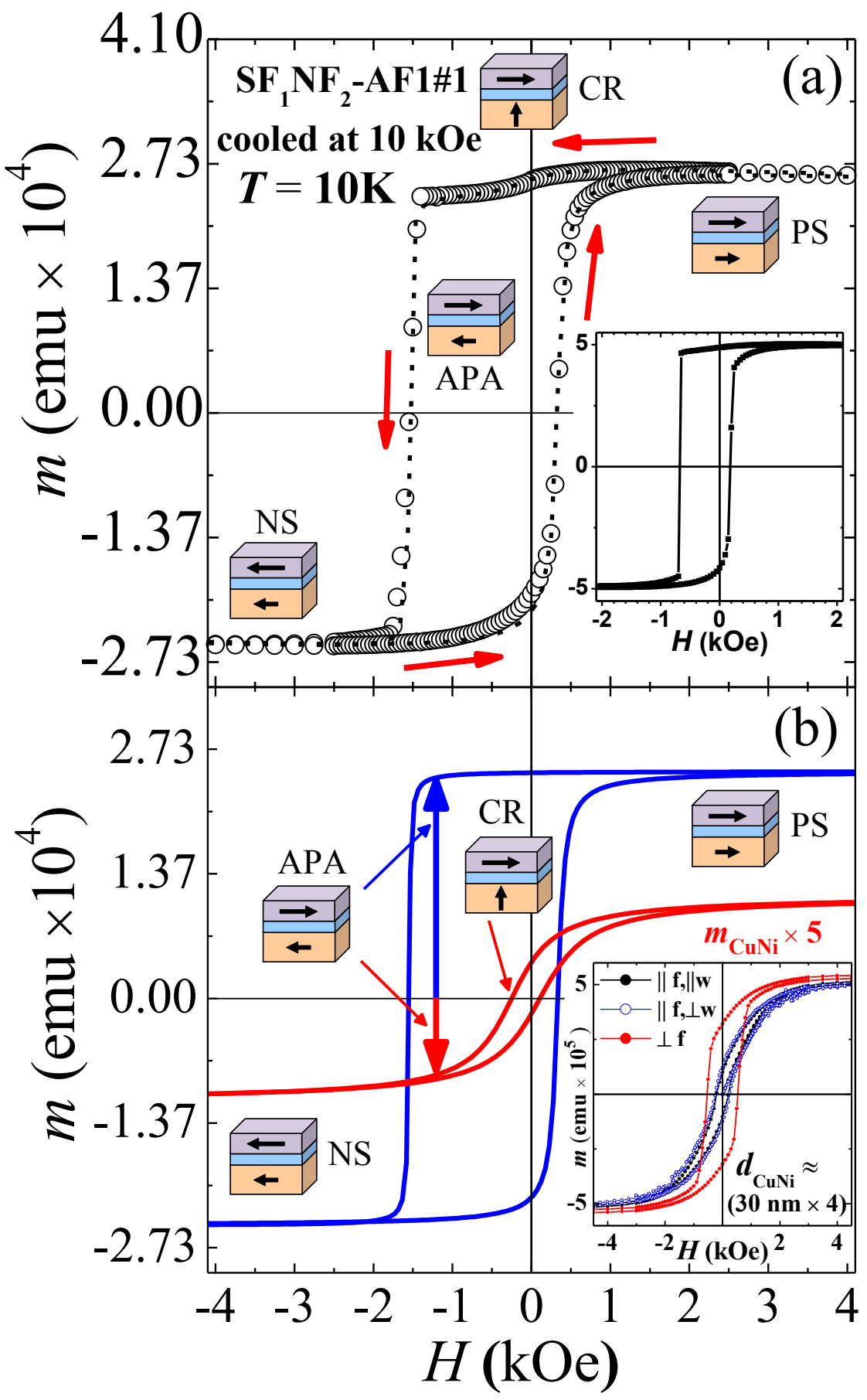

Fig. 2. (a) The magnetic moment, $m$, hysteresis loop (sweep route indicated) of the $Nb/Cu_{41}Ni_{59}/nc\text{-}Nb/Co/CoO_x$ specimen, sample $SF_1NF_2$-AF1#1 ($d_{CuNi} \approx 28$ nm). The dashed line is a modeling according to Ref. [23]. Inset: Hysteresis loop of $SF_1NF_2$-AF1#16 ($d_{CuNi} \approx 11$ nm) adjacent to the sample used for the magnetoresistance measurements below. (b) Modeled components of the hysteresis loop: The blue line is the loop of the cobalt thin film layer, and the red one of the $Cu_{41}Ni_{59}$ (magnified by a factor of 5). Diamagnetic contribution of the Si substrate subtracted. Inset: Hysteresis loops of the reference $(Cu_{41}Ni_{59}/Si)\times4$ sample ($d_{CuNi} \approx 30$ nm), measured at $T = 2$K perpendicular to the film (red ⊥f), in the sample plane perpendicular (blue ||f,⊥w) and parallel (black ||f,||w) to the CuNi layer thickness gradient of the wedge (see above). Abbreviations at the pictograms introduced in the text.

The $R(H)$ measurements in the temperature range of the SC transition for sample $SF_1NF_2$-AF1#17 are presented in Figs. 3 (c, d) and for samples #2 and #24 in Figs. 3 (e, f), respectively. In Figs. 3 (c, e, f) the MR loops were recorded at temperatures fixed close to the middle of the SC transitions at $H = 0$ Oe, while in Fig. 3(d), $T_3 \approx 3.540$ K is close to the end of the transition. For temperatures in the middle of the SC transition, upward MR claws of large magnitude, reaching about 40% of the resistance at ±4 kOe (see Fig. 3(c)), located close to the coercive fields of the $Cu_{41}Ni_{59}$ layer are observed for the samples with thinner CuNi (#17 and #24), whereas broad and flat cusps are found for sample #2 corresponding to the CR-APA range of fields of the loop in Fig. 2. At $T_3 \approx 3.540$ K, sample #17 passes through a sequence of resistive-superconducting-resistive transitions (see Fig. 3(d)) confined to the magnetic configurations in the system. Quantitative comparison of the MR, AMR and $m(H)$ data for the thinner samples (#17 and #24) allows us to identify the spikes positions with the CR magnetic moment configurations of the Co and $Cu_{41}Ni_{59}$ layers.

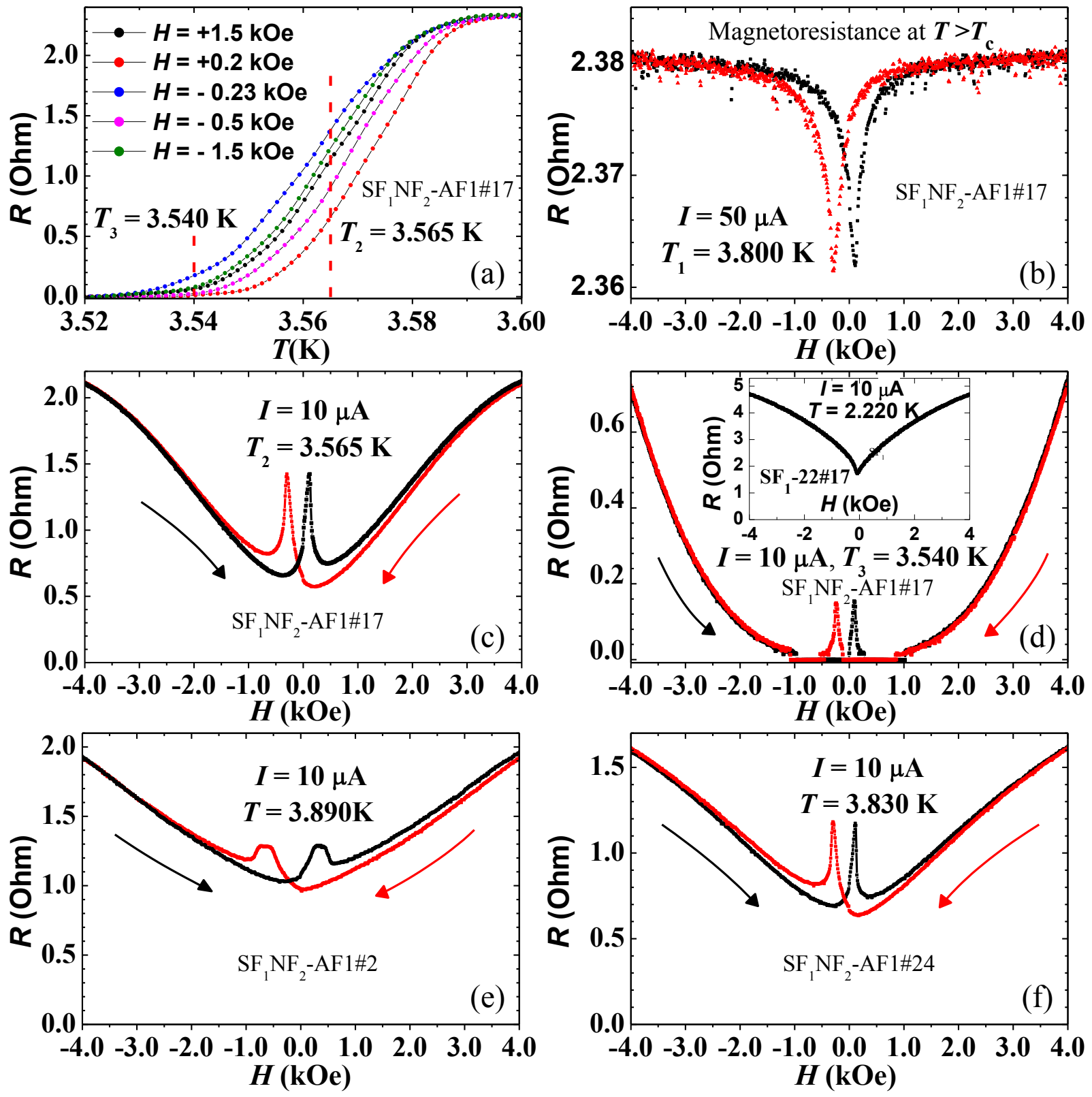

Fig. 3. Experimental results for a $Nb/Cu_{41}Ni_{59}$/nc-Nb/Co/$CoO_x$ spin valve structure ($SF_1NF_2$-AF1 series, $d_{Nb} \approx 12$ nm) measured after cooling down in a field of 30 kOe. (a) SC transition curves at different magnetic fields, sample #17, $d_{CuNi} = 9.9$ nm; (b) Magnetoresistance data recorded well above the SC transition, sample #17; (c) MR curves recorded at $T_2 \approx 3.565$ K, sample #17; (d) MR curves recorded at $T_3 \approx 3.540$ K, sample #17. Inset: MR of the pilot $Nb/Cu_{41}Ni_{59}$/Si-cap bilayer (sample $SF_1$-22#17, on a Si buffer layer one has $d_{Nb} \approx 8$ nm and $d_{CuNi} \approx 15$ nm) measured in the same geometry and sequence as basic sample #17; (e) Magnetoresistance curves of sample #2, $d_{CuNi} = 27.3$ nm; (f) Magnetoresistance curves of sample #24, $d_{CuNi} = 1.8$ nm

Several reasons may generate the unconventional behavior of the SC transition temperature [24, 25]: (i) a magnetic domain structure in the F layers; (ii) Abrikosov vortices induced in the bottom superconducting Nb layer by the $Cu_{41}Ni_{59}$ alloy stray fields at perpendicular alignment of its magnetic moment; (iii) the triplet pairing generation in the spin-valve structure. We excluded the possibility of current dependent quasiparticle accumulation in the NCA state [25,26] because no marked change in MR was observed for currents from 1 μA to 100 μA.

The maze-like domain structure, developed in $Cu_{47}Ni_{53}$ films below the saturation field, has a spatial period of about 100 nm (Ref. [20], Fig. 3), which is much larger than the coherence lengths in our system. The stray field issue is closely related to the vortex-antivortex generation in the superconducting Nb film, and their motion results in a transition temperature reduction [27]. These scenarios were checked with MR measurements of the pilot $Nb/Cu_{41}Ni_{59}$/Si-cap sample in the same geometry and at the mid-transition temperature (see inset in Fig. 3(d)). The same influence of the $Cu_{41}Ni_{59}$ domain structure on superconductivity of the Nb layer is expected as in the $Nb/Cu_{41}Ni_{59}$/nc-Nb/Co/$CoO_x$ structure, however, no upward MR claws were observed in these pilot measurements. Moreover, one would expect the influence of the stray fields to reduce with decreasing the CuNi thickness, however, the upward MR claws for sample #24 with 5 times thinner $Cu_{41}Ni_{59}$ layer (Fig. 3(f)) have a MR magnitude and shape comparable with sample #17 (Fig. 3(c)). These observations controvert (i) and (ii) scenarios.

The experimental findings can be consistently described in the framework of the existing theory of the S/$F_1$/$F_2$ core structure [17,28] (*i.e.* scenario (iii)) The S/$F_1$/$F_2$ core, compared with the $F_1$/S/$F_2$ or $F_1$/S/$F_1$ cores design [29-31], allows not only $T_c$ for the parallel (P) alignment to be lower than for the antiparallel (AP) alignment of the $F_1$ and $F_2$ magnetic moments ($T_c^P < T_c^{AP}$– the "direct" spin–valve effect), but also the opposite ($T_c^{AP} < T_c^P$ – the "inverse" spin–valve effect). Moreover, a non-monotonic dependence of the superconducting transition temperature $T_c$ on the angle between the magnetic moments of the adjacent ferromagnetic layers, $F_1$ and $F_2$, and the "triplet" spin–valve effect was predicted [17], at which $T_c^{TR}$ for the NCA of magnetic moments is the absolute minimum $T_c$, because $T_c^{TR} < \{T_c^{AP}, T_c^P\}$. Recently, the "direct" and "inverse" spin–valve effects were demonstrated in a $CoO_x$/Fe/Cu/Fe/In heterostructure [24]. Below we argue that we could observe the new, "triplet" spin-valve effect in the Nb/$Cu_{41}Ni_{59}$/nc-Nb/Co/$CoO_x$ structure.

Two calculated curves, realizing all regimes mentioned above, are presented in Fig. 4a. The predicted behavior can be compared with our experimental data in Fig. 4b, where $T_c(H)$ taken at the midpoint of the resistive transition are presented. The data recording starts from the positive saturated state at + 4 kOe corresponding to the starting point of the MR measurements in Fig. 3 (c) and (d). After the field polarity change, $T_c(H)$ rapidly drops and reaches the minimal $T_c^{TR}$ at the field close to the negative coercive field (see Fig. 2a). The downward spike in Fig. 4b coincides with the left (red) spikes in Fig. 3c and Fig. 3d and corresponds to the crossed magnetic moments configuration of the $Cu_{41}Ni_{59}$ and the Co layers as indicated by the pictogram. We identify the $T_c$ drop in Fig. 4b with the "triplet" spin-valve effect predicted in Ref. 17.

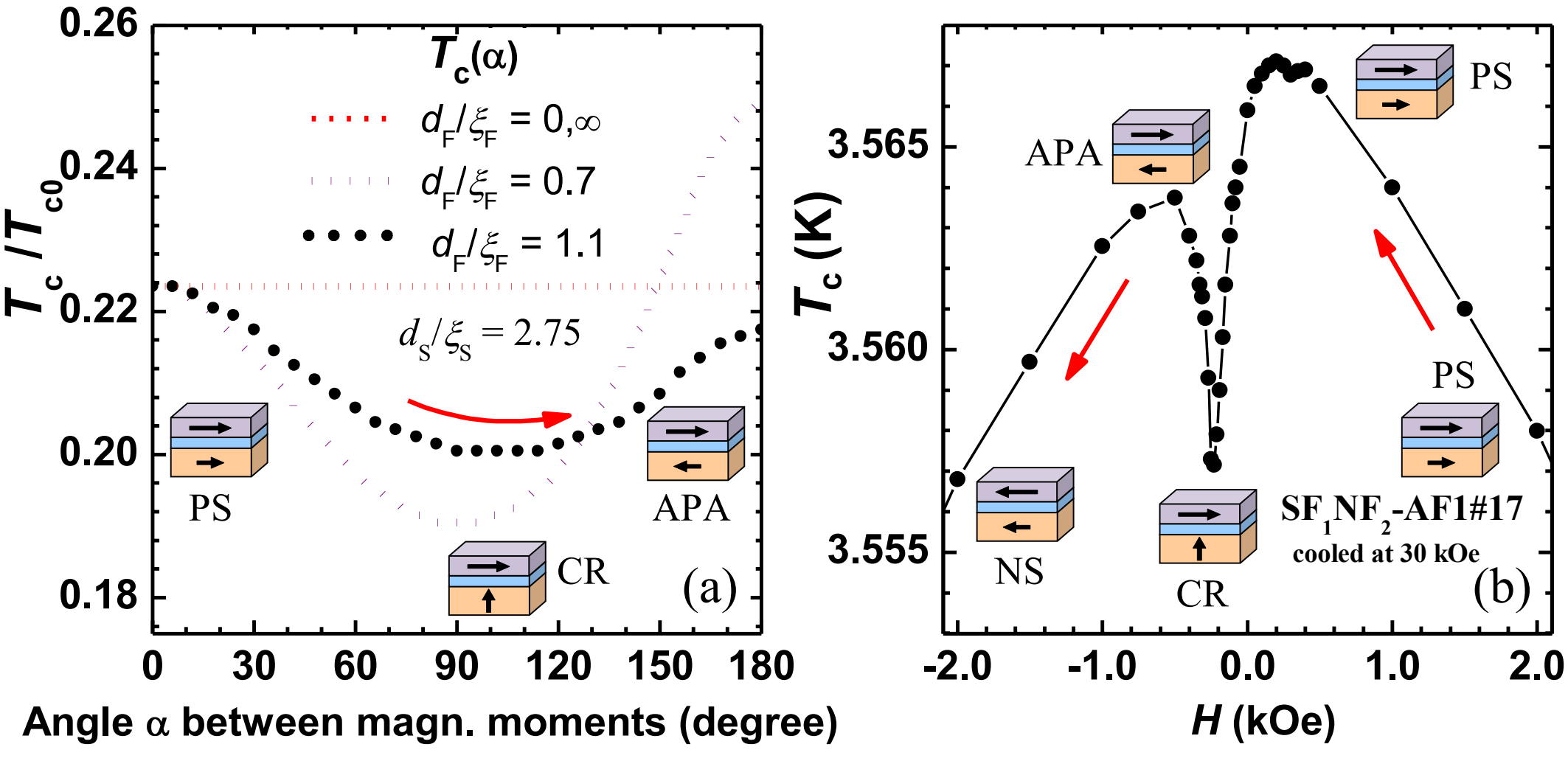

Fig. 4. (a) The angular dependence of the critical temperature $T_c$ according to the model developed in [17]. Here, $T_{c0}$ is the critical temperature, the free-standing S-film would have, (b) Dependence of the critical temperature $T_c$ on the magnetic field, swept along the hysteresis loop shown in Fig. 2. The direction of the field sweep in the panel (b) is shown by red arrows; the corresponding evolution of Tc in the panel (a) occurs along the red arrow. Physically, the negative saturated state is not equal to the initial positive saturated state because the magnetic anisotropy of the system is not uniaxial, but unidirectional (see Fig. 2). Here $d_F$, $d_S$, $\xi_F$, and $\xi_S$ are the thicknesses and the coherence lengths [18] of the ferromagnet $F_1$ and the superconductor respectively

In the theory [17], the two layers of weak ferromagnets are considered with a short electron mean free path. The outer ferromagnetic layer is infinitely thick. Superconductivity in the heterostructure is treated using the Usadel equations [32]. This seems to be not applicable if one of the layers is made of a strong ferromagnet like cobalt. However, the functional layer adjacent to the Nb film, is $Cu_{41}Ni_{59}$, a weak ferromagnetic alloy. The outer ferromagnetic layer in the theory serves as a mixer of the singlet and triplet pairing channels for the adjacent functional layer. Since Co has a very short coherence length $\xi_F$ (Co) ≈ 1.3 nm [33], $d_{Co}/\xi_F(Co) \approx 15.4$ nm/1.3 nm =11.8 (for the thickness $d_{Co}$ see Fig. 1). This is physically infinite as required by the theory.

In summary, we observed experimentally unusual magnetoresistance peaks and sequences of resistive to superconducting and vice versa transitions in the $Nb/Cu_{41}Ni_{59}$/nc-Nb/Co/$CoO_x$ spin-valve heterostructure associated with coercive fields of the $Cu_{41}Ni_{59}$ layer and attributed to a non-collinear magnetic configuration of the ferromagnetic layers in the structure. The superconducting transition temperature shift in a magnetic field and a careful analysis of magnetic configurations in the system allowed us to conclude that we observed experimentally the predicted novel triplet spin-valve effect.

The authors are grateful to V. V. Ryazanov, A. D. Zaikin and R. G. Deminov for stimulating discussions, to D. Vieweg for assistance in magnetic measurements, and to S. Heidemeyer, B. Knoblich, and W. Reiber for assistance in the TEM sample preparation. The work was supported by the Deutsche Forschungsgemeinschaft (DFG) grant No GZ: HO 955/6-1,2 and in part by the Russian Fund for Basic Research (RFBR) under the grants No. 11-02-00848-a (L.R.T.) and 11-02-12065-ofi_m (M.Yu.K.).

*****************************************************************

**References**

[1] F. S. Bergeret, A. F. Volkov, and K. B. Efetov, Phys. Rev. Lett. **86**, 4096 (2001).
[2] A. F. Volkov, F. S. Bergeret, and K. B. Efetov, Phys. Rev. Lett. **90**, 117006 (2003).
[3] F. S. Bergeret, A. F. Volkov, and K. B. Efetov, Rev. Mod. Phys. **77**, 1321 (2005).
[4] M. Eschrig, Phys. Today **64**, 43 (2011).
[5] R. S. Keizer, *et al.,* Nature (London) **439** , 825 (2006).
[6] I. Sosnin, *et al.* Phys. Rev. Lett. **96**, 157002 (2006).
[7] D. Sprungmann, *et al.,* Phys. Rev. B **82**, 060505(R) (2010).
[8] T. S. Khaire, *et al.,* Phys. Rev. Lett. **104** , 137002 (2010).
[9] J. Wang, *et al.*, Nature Phys. **6**, 389 (2010).
[10] J. W. A. Robinson, J. D. S. Witt, and M. G. Blamire, Science **329**, 59 (2010).
[11] M. S. Anwar, *et al.,* Phys. Rev. B **82**, 100501(R) (2010).
[12] T. Yu. Karminskaya and M. Yu. Kupriyanov, Pis'ma v ZhETF **86,** 65 (2007) [JETP Lett. **86**, 61 (2007)].
[13] M. Houzet and A. I. Buzdin, Phys. Rev. B **76**, 060504(R) (2007).
[14] T. Yu. Karminskaya, M. Yu. Kupriyanov, and A. A. Golubov, Pis'ma v ZhETF **87,** 657 (2008) [JETP Lett. **87**, 570 (2008)].
[15] J. Zhu, *et al.,* Phys. Rev. Lett. **105**, 207002 (2010).
[16] J. Y. Gu, J. Kusnadi, and C.-Y. You, Phys. Rev. B **81**, 214435 (2010).
[17] Ya. V. Fominov, *et al.,* Pis'ma v ZhETF **91** , 329 (2010) [JETP Lett. **91**, 308 (2010)].
[18] V. I. Zdravkov, *et al.,* Phys. Rev. B **82** , 054517 (2010).
[19] A. Ruotolo, *et al.,* J. Appl. Phys. **96**, 512 (2004).
[20] I. S. Veshchunov, *et al.,* Pis'ma v ZhETF **88**, 873 (2008) [JETP Lett. **88**, 758 (2008)].
[21] F. Radu and H. Zabel, in *Magnetic Heterostructures*, STMP v. **227** (edited by H. Zabel and S. D. Bader, Springer, Berlin-Heidelberg, 2008), Ch.3.
[22] V. Zdravkov, *et al.,* Phys. Rev. Lett. **97** , 057004 (2006).
[23] A. L. Geiler, *et al.,* J. Appl. Phys. **99**, 08B316 (2006).
[24] P. V. Leksin, *et al,*. Phys. Rev. Lett. **106** , 067005 (2011).

[25] A. Yu. Rusanov, T. E. Golikova, and S. V. Egorov, Pis'ma v ZhETF **87**, 204 (2008) [JETP Lett. **87**, 175 (2008)]
[26] A. Yu. Rusanov, S. Habraken, and J. Aarts, Phys. Rev. B **73,** 060505(R) (2006)
[27] I. S. Burmistrov and N. M. Chtchelkatchev, Phys. Rev. B **72**, 144520 (2005).
[28] S. Oh, D. Youm, and M. R. Beasley, Appl. Phys. Lett. **71** , 2376 (1997).
[29] L. R. Tagirov, Phys. Rev. Lett. **83**, 2058 (1999).
[30] Ya. V. Fominov, A. A. Golubov, and M. Yu. Kupriyanov, Pis'ma v ZhETF **77**, 609 (2003) [JETP Lett. **77**, 510 (2003)].
[31] J. Kehrle, *et al.,* Ann. der Physik (Berlin) **524**, 37 (2012).
[32] K. D. Usadel, Phys. Rev. Lett. **25**, 507 (1970).
[33] Jun-Jih Liang, *et al.,* J. Appl. Phys. **92** , 2624 (2002).